\documentclass{ws-p8-50x6-00}

\begin{document}

\title{Monopole potential and confining strings in the 
(2+1)-dimensional Georgi-Glashow model}

\author{Dmitri Antonov}

\address{INFN-Sezione di Pisa, Universit\'a degli studi di Pisa,
Dipartimento di Fisica, Via Buonarroti, 2 - Ed. B - I-56127 Pisa, 
Italy\\
E-mail: antonov@df.unipi.it}


\maketitle

\abstracts{Confining strings are investigated in the 
(2+1)D Georgi-Glashow model.
This is done
in the limit when the electric coupling constant is much larger 
than the square root of the mass of the 
Higgs field, but much smaller
than the vacuum expectation value of this field.  
The modification of the Debye mass of the dual photon 
with respect to the case when it is considered to be negligibly small
compared to the Higgs mass, is found. Analogous modifications 
of the potential of monopole 
densities and string coupling constants are found as well.}

(2+1)D Georgi-Glashow model is known to be the famous example of a 
theory allowing for an analytical description of confinement~\cite{1}.
The confinement mechanism in this model is based on the presence  
of 't Hooft-Polyakov monopoles~\cite{2}, whose stochastic fluxes 
through the contour of the Wilson loop provide the area law.
In this talk, we shall discuss string representation 
of the Wilson loop in this model for the case when, contrary to the 
compact-QED limit,  
the Higgs mass is not considered to be infinitely large {\it w.r.t.} the 
Debye mass of the dual photon.

The Euclidean action of the (2+1)D Georgi-Glashow model has the 
following form

\begin{equation}
\label{GG}
S=\int d^3x\left[\frac{1}{4g^2}\left(F_{\mu\nu}^a\right)^2+
\frac12\left(D_\mu\Phi^a\right)^2+\frac{\lambda}{4}\left(
\left(\Phi^a\right)^2-\eta^2\right)^2\right],
\end{equation}
where the Higgs field $\Phi^a$ transforms by the adjoint representation, 
and $D_\mu\Phi^a\equiv\partial_\mu\Phi^a+\varepsilon^{abc}A_\mu^b
\Phi^c$.
In the one-loop approximation, the 
partition function of this theory reads~\cite{dietz}

$$
{\cal Z}=1+\sum\limits_{N=1}^{\infty}\frac{\zeta^N}{N!}
\left[
\prod\limits_{i=1}^{N}\int d^3z_i\sum\limits_{q_i=\pm 1}^{}\right]
\times
$$

\begin{equation}
\label{1}
\times\exp\left\{-\frac{g_m^2}{2}\left[\int d^3xd^3y\rho_{\rm gas}({\bf x})
D_0({\bf x}-{\bf y})
\rho_{\rm gas}({\bf y})-
\sum\limits_{{a,b=1\atop a\ne b}}^{N}
D_m({\bf z}_a-{\bf z}_b)
\right]\right\}.
\end{equation}
Here, $g_m$ is the magnetic coupling constant of dimensionality 
$[{\rm length}]^{1/2}$ related to the electric one $g$
according to the equation $gg_m=4\pi$, $\rho_{\rm gas}({\bf x})=
\sum\limits_{a=1}^{N}q_a\delta\left({\bf x}-{\bf z}_a\right)$
is the density of monopole gas with $q_a$'s standing for the 
monopole charges in the units of $g_m$. Next, in Eq.~(\ref{1}), 
$m=\eta\sqrt{2\lambda}$ is the mass of the Higgs boson and 

\begin{equation}
\label{zeta}
\zeta=\frac{m_W^{7/2}}{g}\delta\left(\frac{\lambda}{g^2}\right)
{\rm e}^{-(4\pi/g^2)m_W\epsilon\left(\lambda/g^2\right)}
\end{equation}
is the statistical weight of a single monopole (else called fugacity)
with $m_W=g\eta$ being the mass of the $W$-boson.
Here, $\epsilon$ is a slowly varying function equal to unity at the 
origin [{\it i.e.} in the Bogomolny-Prasad-Sommerfield 
(BPS) limit]~\cite{bps} and $1.787\ldots$ at infinity~\cite{kirk}, whereas 
the function $\delta$ is determined by the loop corrections.
Finally, in Eq.~(\ref{1}), $D_0({\bf x})\equiv1/(4\pi|{\bf x}|)$ 
is the Coulomb propagator, and 
$D_m({\bf x})\equiv{\rm e}^{-m|{\bf x}|}/(4\pi|{\bf x}|)$
is the propagator of the Higgs boson.

Notice that as it follows from Eq.~(\ref{1}), in the BPS limit, 
the interaction of two monopoles doubles for opposite and vanishes 
for equal charges. As far as the opposite limit, $m\to\infty$, is 
concerned, we apparently arrive  
there at the standard compact-QED result~\cite{1}.

The effective field theory describing the grand canonical 
partition function~(\ref{1}) can easily be obtained and reads~\cite{dietz}

$$
{\cal Z}=\int {\cal D}\chi{\cal D}\psi\times
$$

\begin{equation}
\label{2}
\times\exp\left\{-\int d^3x\left[
\frac12(\nabla\chi)^2+\frac12(\nabla\psi)^2+\frac{m^2}{2}\psi^2-
2\zeta{\rm e}^{g_m\psi}\cos(g_m\chi)\right]\right\},
\end{equation}
where $\chi$ is the dual photon field, whereas the field $\psi$ is an
additional one. The latter field can be integrated out in the 
limit $g\gg\sqrt{m}$,
when the exponent in the last term 
on the R.H.S. of Eq.~(\ref{2})
can be shown~\cite{ND} 
to be approximated by the terms not higher than the linear one.

In such a limit, 
Gaussian integration over the field $\psi$ yields

$$
{\cal Z}=\int {\cal D}\chi\exp\left\{-\int d^3x\left[
\frac12(\nabla\chi)^2-
2\zeta\cos(g_m\chi)\right]+\right.
$$

\begin{equation}
\label{3}
\left.+2(g_m\zeta)^2\int d^3xd^3y\cos(g_m\chi
({\bf x}))D_m({\bf x}-{\bf y})\cos(g_m\chi({\bf y}))\right\}.
\end{equation}
The last term here represents the correction to the standard
result~\cite{1}. It stems from the fact that the mass of the 
Higgs field was considered to be not infinitely large compared to the
standard Debye mass of the dual photon, $m_D=g_m\sqrt{2\zeta}$.
The respective correction to $m_D$ is positive, and the square of the 
full mass reads: 

\begin{equation}
\label{M}
M^2=m_D^2\left(1+\frac{m_D^2}{m^2}\right).
\end{equation} 
Clearly, this result is valid at $m_D\ll m$ and reproduces $m_D^2$ 
in the limit $m\to\infty$.

Another relation between the dimensionful parameters in the 
model~(\ref{GG}), we shall adapt for our analysis, is 
$g\ll\eta$.
[Clearly, this inequality parallels the requirement that $\eta$ should
be large enough to ensure the spontaneous symmetry breaking from $SU(2)$
to $U(1)$.] In particular, from this relation and the
inequality $g\gg\sqrt{m}$ 
we immediately obtain:

\begin{equation}
\label{lambda}
\frac{\lambda}{g^2}\sim\left(\frac{m}{m_W}\right)^2\ll\left(\frac{g}{\eta}
\right)^2\ll 1.
\end{equation}
This means that we are working in the regime of the Georgi-Glashow 
model close to the BPS limit.

Note further that in the limit $g\gg\sqrt{m}$, 
the dilute gas approximation holds perfectly. 
Indeed, this approximation implies
that the mean distance between monopoles, equal to $\zeta^{-1/3}$, 
should be much larger than the inverse mass of the $W$-boson. By virtue of 
Eq.~(\ref{zeta}) and the fact that the function $\epsilon$ is of the order
of unity, we obtain that this 
requirement is equivalent to the following one: 

\begin{equation}
\label{small}
\sqrt{\frac{\eta}{g}}\delta\left(\frac{\lambda}{g^2}\right)
{\rm e}^{-4\pi\eta/g}\ll 1.
\end{equation}
Although at $\lambda\ll g^2$ [{\it cf.} Eq.~(\ref{lambda})],
the function $\delta$ grows, the speed of this growth 
is so that at $g\ll\eta$,
the L.H.S. of Eq.~(\ref{small}) remains exponentially small~\cite{ks}.
Another consequence of this fact is that in the regime
of the Georgi-Glashow model under discussion, the Debye mass 
of the dual photon, $m_D$, remains exponentially small as well.
In particular, the inequality $m_D\ll m$, under which the full mass~(\ref{M})
was derived, holds due to this smallness. Also, due to the same reason,
the mean field approximation, under which the effective field
theory~(\ref{2}) is applicable, remains valid as well with the 
exponential accuracy. Indeed, 
this approximation implies that in the Debye 
volume~\footnote{In this discussion, the difference
between $m_D$ and $M$ is unimportant.} 
$m_D^{-3}$ there should contain 
many particles~\cite{1}. Since the average density of monopoles
is equal to $2\zeta$ [which can be seen either by calculating it directly
according to the formula $V^{-1}\partial\ln{\cal Z}/\partial\ln\zeta$,   
applied to Eq.~(\ref{3}) at $m_D\ll m$, or 
from the remark following after Eq.~(\ref{7}) below], 
we arrive at the requirement $\zeta m_D^{-3}\gg 1$. 
Substituting the above-obtained value for $m_D$, we see that the 
criterion of applicability of the mean field approximation 
reads $g^3\gg\sqrt{\zeta}$. Owing to the above-discussed exponential
smallness of $\zeta$, this inequality is satisfied.

One can now derive the potential of (dynamical) monopole densities 
corresponding to the partition function~(\ref{3}). Referring the 
reader for the details to the original paper~\cite{ND}, we present here 
the final expression for the partition 
function~(\ref{3}) in terms of the integral over these densities:

\begin{equation}
\label{5}
{\cal Z}=\int {\cal D}\rho\exp\left\{-\left[
\frac{g_m^2}{2}\int d^3xd^3y\rho({\bf x})D_0({\bf x}-
{\bf y})\rho({\bf y})
+V[\rho]\right]\right\}.
\end{equation}
The monopole potential $V[\rho]$ here reads

$$
V[\rho]=\int d^3x\left[\rho{\,}{\rm arcsinh}
\varrho-
2\zeta\sqrt{1+\varrho^2}\right]-
$$

\begin{equation}
\label{6}
-2(g_m\zeta)^2
\int d^3xd^3y\sqrt{1+\varrho^2({\bf x})}
D_m({\bf x}-{\bf y})\sqrt{1+\varrho^2({\bf y})},
\end{equation}
where $\varrho\equiv\rho/(2\zeta)$.
The last term on the R.H.S. of this equation is again a leading
$(m_D/m)$-correction to the respective $(m=\infty)$-expression.
In the dilute gas approximation, $|\rho|\ll\zeta$, 
Eq.~(\ref{6}) becomes a simple quadratic functional:

$$V[\rho]\to\frac12\left[\frac{1}{2\zeta}-\left(\frac{g_m}{m}
\right)^2\right]\int d^3x\rho^2\simeq\frac{g_m^2}{2M^2}\int d^3x\rho^2,$$
where the last equality is implied within the leading $(m_D/m)$-approximation
adapted.
This leads to the simple expression for the generating functional
of correlators of the monopole densities within these two approximations:

$$
{\cal Z}[j]\equiv
\int {\cal D}\rho\times$$

$$\times\exp\left\{-\left[
\frac{g_m^2}{2}\int d^3xd^3y\rho({\bf x})D_0({\bf x}-
{\bf y})\rho({\bf y})
+\frac{g_m^2}{2M^2}\int d^3x\rho^2+
\int d^3xj\rho\right]\right\}=$$

\begin{equation}
\label{7}
=\exp\left[-\frac{M^2}{2g_m^2}\int d^3xd^3yj({\bf x})
j({\bf y})\partial^2 D_M({\bf x}-{\bf y})\right].
\end{equation}
[Sending for a while $m_D$ to zero (since it is exponentially small), 
we get from Eq.~(\ref{7}): 
$\left<\rho({\bf x})\rho(0)\right>=2\zeta\delta({\bf x})$. 
This means that with the exponential accuracy 
the average density of monopoles is equal to $2\zeta$, which can also be 
seen directly from the $(|\rho|\ll\zeta)$-limit of Eq.~(\ref{6}).]
In particular, the Wilson loop reads:

$$
\left<W({\cal C})\right>=\left<W({\cal C})\right>_{\rm free}{\cal Z}[i\eta]=
$$

\begin{equation}
\label{8}
=\exp\left\{-\frac{g^2}{2}\left[\frac{M^2}{2}\int\limits_{\Sigma}^{} 
d\sigma_{\mu\nu}({\bf x})
\int\limits_{\Sigma}^{} d\sigma_{\mu\nu}({\bf y})+\oint\limits_{{\cal C}}^{}
dx_\mu \oint\limits_{{\cal C}}^{}dy_\mu\right]D_M({\bf x}-{\bf y})
\right\}.
\end{equation}
This equation can straightforwardly be derived by making use of the 
formula 

$$\partial_\mu\eta({\bf x})=2\pi\varepsilon_{\mu\nu\lambda}
\left[2\partial_\nu^x\oint\limits_{{\cal C}}^{}dy_\lambda
D_0({\bf x}-{\bf y})-\int\limits_{\Sigma}^{}d\sigma_{\nu\lambda}({\bf y})
\delta({\bf x}-{\bf y})\right].$$
Here, $\eta({\bf x})=2\pi\varepsilon_{\mu\nu\lambda}\partial_\mu^x
\int\limits_{\Sigma}^{}d\sigma_{\nu\lambda}({\bf y})D_0({\bf x}-{\bf y})$
is the solid angle under which an arbitrary surface $\Sigma$ 
spanned by the contour ${\cal C}$ shows up to the observer located at the 
point ${\bf x}$. Also, in Eq.~(\ref{8}), 
$\left<W({\cal C})\right>_{\rm free}=\exp\left[-\frac{g^2}{2}
\oint\limits_{{\cal C}}^{}dx_\mu \oint\limits_{{\cal C}}^{}dy_\mu 
D_0({\bf x}-{\bf y})\right]$ is the contribution of the free photons
to the Wilson loop. The explicit $\Sigma$-dependence of the R.H.S.
of Eq.~(\ref{8}) appearing in the dilute gas approximation 
becomes eliminated by the summation over branches of the 
arcsinh-function in the full monopole action~(\ref{5})-(\ref{6}).
This is the main principle of correspondence between fields
strings, proposed for compact QED in Ref.~\cite{cs} in the 
language of the Kalb-Ramond field $h_{\mu\nu}$, 
$\varepsilon_{\mu\nu\lambda}\partial_\mu h_{\nu\lambda}\propto\rho$.

As far as the string tension and the inverse coupling constant of
the rigidity term~\cite{rid} are concerned, those can be evaluated 
upon the derivative expansion of the $\Sigma$-dependent part of 
Eq.~(\ref{8}). 
By virtue of the general formulae from Ref.~\cite{aes} we obtain 

\begin{equation}
\label{sigalfa}
\sigma=4\pi g^2M~~ {\rm and}~~
\alpha^{-1}=-\frac{\pi g^2}{2M},
\end{equation}
respectively. Clearly, both of 
these quantities
represent the modifications of the standard 
ones, corresponding to the limit when $m$ is considered to be 
infinitely large {\it w.r.t.} $m_D$. The standard expressions
follow from Eq.~(\ref{sigalfa}) upon the substitution 
into this equation $M\to m_D$.

\section*{Acknowledgments}
The author is grateful to Prof. A. Di Giacomo for useful discussions
and to INFN for the financial support. He was also partially supported
by the INTAS grant Open Call 2000, project No. 110.
And last but not least, the author acknowledges the organizers
of the Sixth
Workshop on Nonperturbative QCD (American University of Paris, 
Paris, France, June 2001)
for an opportunity to present these results in a very stimulating 
atmosphere.

\end{document}